%% file: conference_101719.tex
\documentclass[conference]{IEEEtran}
\IEEEoverridecommandlockouts
\usepackage{cite}
\usepackage{amsmath,amssymb,amsfonts}
\usepackage{algorithmic}
\usepackage{graphicx}
\usepackage{textcomp}
\usepackage{xcolor}
\def\BibTeX{{\rm B\kern-.05em{\sc i\kern-.025em b}\kern-.08em
    T\kern-.1667em\lower.7ex\hbox{E}\kern-.125emX}}

\input{header}

\begin{document}
\bstctlcite{IEEEexample:BSTcontrol}

\title{An Open Source Power System Simulator in Python for Efficient Prototyping of WAMPAC Applications
}

\author{\IEEEauthorblockN{Hallvar Haugdal, Kjetil Uhlen}
\IEEEauthorblockA{Department of Electrical Power Engineering\\
Norwegian University of Science and Technology\\
Trondheim, Norway\\
hallvar.haugdal@ntnu.no, kjetil.uhlen@ntnu.no}
}

\maketitle

\input{main}

\balance
\bibliographystyle{IEEEtran}
\bibliography{references}

\end{document}

%% file: header.tex
\usepackage[utf8]{inputenc}
\usepackage{amsmath, amssymb, bm}
\usepackage{physics}
\usepackage{fancyvrb}
\usepackage{arydshln} 
\usepackage{balance}
\makeatletter
 \renewcommand*\env@matrix[1][*\c@MaxMatrixCols c]{%
    \hskip -\arraycolsep
    \let\@ifnextchar\new@ifnextchar
 \array{#1}}
\makeatother

\UseRawInputEncoding

%% file: main.tex




\begin{abstract}
An open source software package for performing dynamic RMS simulation of small to medium-sized power systems is presented, written entirely in the Python programming language. The main objective is to facilitate fast prototyping of new wide area monitoring, control and protection applications for the future power system by enabling seamless integration with other tools available for Python in the open source community, e.g. for signal processing, artificial intelligence, communication protocols etc. The focus is thus transparency and expandability rather than computational efficiency and performance.

The main purpose of this paper, besides presenting the code and some results, is to share interesting experiences with the power system community, and thus stimulate wider use and further development. Two interesting conclusions at the current stage of development are as follows:

First, the simulation code is fast enough to emulate real-time simulation for small and medium-size grids with a time step of 5 ms, and allows for interactive feedback from the user during the simulation. Second, the simulation code can be uploaded to an online Python interpreter, edited, run and shared with anyone with a compatible internet browser. Based on this, we believe that the presented simulation code could be a valuable tool, both for researchers in early stages of prototyping real-time applications, and in the educational setting, for students developing intuition for concepts and phenomena through real-time interaction with a running power system model.

\end{abstract}

\begin{IEEEkeywords}
Dynamic RMS simulation, wide area monitoring, control and protection, real-time simulation, Python
\end{IEEEkeywords}

\section{Introduction}

Python ranks high among the worlds most popular programming languages. Although very slow compared to many other alternatives (e.g. C++), it has the advantage of being easy to learn, read and write. Development of new applications is therefore generally considered to be much easier and faster in Python. Furthermore, a rich library of open source packages is available, making it straightforward to make use of artificial intelligence, communication protocols, powerful visualizations and more.

In this paper, we present an open source package for performing dynamic RMS-simulations, written entirely in Python. With this, we seek to contribute with a highly transparent, easily modifiable and expandable power system simulator that allows tight integration with powerful tools that already exist in the open source community. We believe that open solutions are essential for developing the wide area monitoring, control and protection applications of the future power system.

The initial motivation for development of such a package arose during research towards a controller for damping of power oscillations. The particular study required a non-linear dynamic power system simulator with detailed generator models, capable of incorporating signal processing techniques like Kalman Filters in the simulation loop, as well as the ability to perform eigenvalue analysis on linearizations of the model. This was found to be surprisingly difficult.

Using Python as a scripting language is standard functionality in many commercial tools, for instance DIgSILENT PowerFactory, and is well described in the documentation of the software. Including Python in the dynamic simulation loop is much more difficult, and is not standard functionality. Interfacing Python with dynamic simulations in PowerFactory can be achieved with the repository described in \cite{Lopez2019}. This functions by making PowerFactory call a dll extension a number of times during each simulation step. The dll file, compiled from C code, calls Python. This potentially expands the functionality of PowerFactory a great deal, but might to some appear complicated to use, for instance requiring some knowledge on compilation C code.

PSS/E is very well interfaced with Python through the psspy module. However, small signal stability analysis requires separate modules to be installed. These modules are not that well interfaced with Python, and potentially complicates the workflow. Also, the full, linearized system matrix, which might be required in some applications, is not directly available.

Among open source alternatives for dynamic power system simulations, we have DPsim\cite{Mirz2019}, which is a simulator written in highly efficient C++ code, specifically developed for real-time- and co-simulation. 
ANDES \cite{Cui2020} is a Python software for symbolic modelling and numerical analysis of power systems. 
The Open-Instance Power System Library (OpenIPSL) \cite{Vanfretti2016}, \cite{Baudette2018}, written in the Modellica language, specifically targets unambiguous model sharing among utilities and researchers.

The above mentioned tools are powerful enough to simulate systems with thousands of buses. This is necessary for these tools to be able to perform simulations on detailed models of real transmission grids. However, for research or educational purposes, smaller aggregated systems with tens to hundreds of buses are often used. Frequently used test systems include the Kundur Two-Area System, the IEEE 9, 14, 39 and 68-bus systems and the Nordic 32 and 44-test systems. Systems of this size can be simulated entirely in Python, without strictly requiring sophisticated performance-boosting techniques that in some cases compromise transparency and expandability.

Having developed the simulation code, the encouraging experience obtained this far regarding performance and ease of use motivates us to present the code in this paper.

The entire code is developed in Python, where the core functionality relies only on standard packages like NumPy \cite{Harris2020} and SciPy \cite{Virtanen2020}. This comes with the advantage that the software is cross platform-compatible and easily deployable, and does not require compilation. Further, the software can easily be uploaded to an online Python interpreter and shared, edited and run in the cloud, which could represent a major benefit for reproducibilty of research.

The code is available in a repository called DynPSSimPy \cite{DynPSSimPy} (Dynamic Power System Simulator in Python) on GitHub, released under the GPLv3 license. By following links provided on the GitHub-page, simple examples can be launched online (using Binder \cite{Jupyter2018}) to demonstrate some of the basic functionality of the software.

Finally, it should be mentioned that the package is in its very early stages of development at this point, with a limited number of dynamic models and somewhat limited functionality in some areas.

The rest of this paper is structured as follows: Section \ref{sec:CoreFunctionality} goes into detail on the core functionality; in Section \ref{sec:Validation} simulation results are validated against commercial simulation software; in  Section \ref{sec:RealTimeSimulation} we elaborate on the real-time simulation functionality; finally, Sections \ref{sec:Discussion} and \ref{sec:Conclusion} contain discussion and conclusions.


\section{Core functionality}
\label{sec:CoreFunctionality}

In dynamic analysis of large scale power systems, the ac electrical variables (voltages and currents) are usually represented as phasors when the focus is on electromechanical dynamics and primary control. The relevant models applied are referred to as RMS-models. The fundamental problem that needs to be solved by the RMS-model simulator is the integration of the differential equations of all the dynamic models in the system, while at  the same time making sure that the algebraic equations representing the network are satisfied.

\subsection{Dynamic RMS-simulation}
The Differential Algebraic Equations (DAE) describing the dynamics of the system can be written on the form
\begin{equation}
    \begin{split}
        \dot{\mathbf{x}} = f(\mathbf{x}, \mathbf{y})\\
        0 = g(\mathbf{x}, \mathbf{y})
    \end{split}
\end{equation}
The function $f$ describes the differential equations for all states, while $g$ describes the algebraic equations representing the network equations. $\mathbf{x}$ is the vector of state variables, while $\mathbf{y}$ is the vector of algebraic variables.

With the dynamic models implemented this far, the algebraic variables are constituted solely by the bus voltages of the simulated system. The set of algebraic equations is linear, and can be written on the form
\begin{equation}
    \mathbf{Y}\mathbf{V} = \mathbf{I}_{inj}(\mathbf{x})
    \label{eq:algebraic_equations_1}
\end{equation}
where $\mathbf{Y}$ is the admittance matrix, $\mathbf{I}_{inj}(\mathbf{x})$ is the vector of current injections and $\mathbf{V}=\mathbf{y}$ the vector of bus voltages.

Solving the algebraic equations is very efficient with the system sizes considered this far (up to 45 buses), allowing us to convert the DAE system into a system of Ordinary Differential Equations (ODE) without significant deterioration of performance.

First, we rewrite the algebraic equations as

\begin{equation}
    \mathbf{y} = \mathbf{V} = \mathbf{Y}^{-1}\mathbf{I}_{inj}(\mathbf{x}) = h(\mathbf{x})
    \label{eq:algebraic_equations_2}
\end{equation}
(In practice, of course, the admittance matrix is not inverted for each time step, but rather it is solved by an efficient algorithm for sparse systems of linear equations, for instance using the function scipy.sparse.linalg.spsolve.)

Further, we eliminate the algebraic variables:
\begin{equation}
    \dot{\mathbf{x}} = f(\mathbf{x}, h(\mathbf{x})) = f(\mathbf{x})
    \label{eq:state_derivative_function}
\end{equation}
This system of ODEs can now be integrated with any suitable integration method, for instance the Runge Kutta-implementation found in scipy.integrate.RK45 \cite{Virtanen2020}.


For real-time simulation, it makes sense to use integration methods with fewer evaluations of the ODE function per step taken by the solver. This allows the step size to be decreased, which in turn results in more accurate representation of continuously changing inputs. The Euler- or Modified Euler methods, requiring one or more evaluations, are thus better alternatives than the fourth order Runge Kutta-method, which requires four evaluations per time step.

\subsection{Initialization of dynamic simulation}
To make sure that the system stays on the initial operating point after the simulation started (assuming no disturbances are applied), the system must be initialized such that all time-derivatives are equal to zero. Essentially, this means determining the state vector $\mathbf{x}_0$ that gives 

\begin{equation}
    0 = f(\mathbf{x}_0)
\end{equation}

Initialization of the simulation can be summarized as follows:
\begin{enumerate}
    \item Perform a Newton-Rhapson power flow calculation to determine the bus voltages and active and reactive power injections at each bus.
    \item From the power flow result, determine the terminal voltage and the active and reactive power produced by each generator.
    \item Perform Kron-reduction of the dynamic admittance matrix. All other buses than the generator buses can be eliminated. However, other buses that should not be eliminated can also be specified. For instance, line outages require less computation effort if the buses of the line in question are present in the reduced system, which might be interesting when performing real-time simulation.
    \item The number of states is determined, and the ordering of the state vector is determined. According to the dynamic model definitions in the model library and the number of instances of each model, the appropriate number of states is added. (For instance, for 10 generators with 6 states each, 60 states is added in the state vector.)
    \item The state vector that makes all the differential equations equal to zero is found. This is done by solving the algebraic equations resulting from setting the differential equations equal to zero.
\end{enumerate}

\subsection{Modal Analysis}
Modal analysis is performed by numerical linearization of the system of ODEs. We find the linearized system matrix as the Jacobian of the ODE function, as follows:
\begin{equation}
    \mathbf{A} = \frac{\partial f(\mathbf{x})}{\partial\mathbf{x}}
\end{equation}

Finally, we perform an eigendecomposition of the system matrix to get the eigenvalues and eigenvectors of the system at this operating point (for instance using the function numpy.linalg.eig).

\subsection{Dynamic models}
\label{sec:DynamicModels}
At present, one generator model, one AVR, one GOV and one PSS model is implemented. Generator controls can be applied on arbitrary units with varying parameters. Furthermore, the software is written such that implementation of additional models is possible.

The formulation of the differential equations representing the synchronous machine is based on the formulation in \cite{Machowski2008}:
\begin{equation}
\begin{split}  
    M\Delta\dot{\omega} =& P_m - P_e\\
    \dot{\delta} =& \Delta \omega\\
    T_{d0}^{'} \dot{E_q^{'}} =& E_f - E_q^{'} - I_d ( X_d - X_d^{'} )\\
    T_{q0}^{'} \dot{E_d^{'}} =& - E_d^{'} + I_q ( X_q - X_q^{'} )\\
    T_{d0}^{''} \dot{E_q^{''}} =& E_q^{'} - E_q^{''} - I_d ( X_d^{'} - X_d^{''} )\\
    T_{q0}^{''} \dot{E_d^{''}} =& E_d^{'} - E_d^{''} + I_q ( X_q^{'} - X_q^{''} )\\
\end{split}
\end{equation}

Subtransient saliency is neglected, i.e. we require $X_d^{''} = X_q^{''}$. Interfacing each generator with the grid is achieved using the following equations:
\begin{equation}
    \begin{bmatrix}
        E_d^{''} \\ E_q^{''}
    \end{bmatrix}
    =
    \begin{bmatrix}
        V_d \\ V_q
    \end{bmatrix}
    +
    \begin{bmatrix}
        R & X_q^{''} \\ X_q^{''} & R
    \end{bmatrix}
    \begin{bmatrix}
        I_d \\ I_q
    \end{bmatrix}
\end{equation}

Using that $X_d^{''} = X_q^{''} = X^{''}$, this can be simplified to
\begin{equation}
    \begin{split}
    E^{''}&= E_d^{''} + jE_q^{''}\\
    &=(V_d + jV_q) + ( R +jX^{''})(I_d+jI_q)\\
    &= V + ( R +jX^{''})I
    \end{split}
\end{equation}

Further, we establish the Northon Equivalent of the synchronous machine, where the current source is given by
\begin{equation}
    I_{no} = E^{''}/X^{''}
\end{equation}
and the shunt impedance by
\begin{equation}
    Z_{no} = R + jX^{''}
\end{equation}

The impedances $Z_{no}$ of each of the synchronous machines are included in the admittance matrix used during dynamic simulation, $\mathbf{Y}$ in \eqref{eq:algebraic_equations_1}, appearing as contributions on the diagonal entries. The currents $I_{no}$ appear as contributions to the current injection vector $\mathbf{I}_{inj}(\mathbf{x})$.

Finally, the electrical power is given by

\begin{equation}
    P_e = E_d^{''}I_d + E_q^{''}I_q
\end{equation}

For generator controls, the standard models SEXS for AVR, TGOV1 for turbine/governor and STAB1 for PSS are implemented.

\subsection{Implementation}
Each dynamic model is defined as a class in Python. Within each class, the function for calculating the state derivatives of the dynamic model is defined. Functions for initialization and computing current injections are also implemented for models where this is required.

Evaluating the complete state derivative vector in \eqref{eq:state_derivative_function} is achieved by looping through the state derivative functions of all the individual dynamic models included in the power system model.

For the implemented models described above, the state derivative functions make use of vectorization, which significantly improves the computational efficiency. A further increase in computational efficiency can be achieved by leveraging Just-In-Time compilation of the derivative functions of each model, available in Python using Numba \cite{Lam2015}. This introduces some overhead, but speeds up the state derivative functions significantly once the compilation is done.

\section{Validation}
\label{sec:Validation}

\begin{figure}
    \centering
    \includegraphics{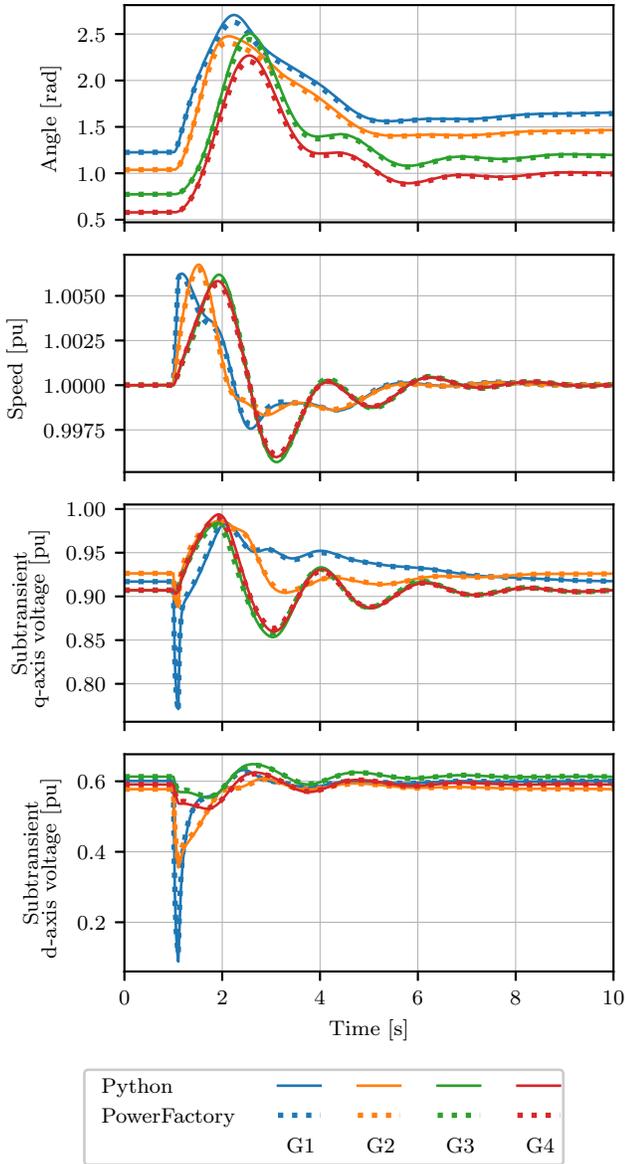}
    \caption{The results from simulating the Kundur Two-Area System in Python are shown alongside corresponding results from a model with identical parameters simulated in PowerFactory. The transient d- and q-axis voltages are not shown, since these are not directly available from the PowerFactory simulation (since in PowerFactory the dynamic generator equations are formulated in terms of fluxes instead of voltages). The results are not identical, but very similar.}
    \label{fig:K2A-Validation}
\end{figure}

The simulator is validated by comparing results from the Kundur Two-Area System, obtained using the presented Python package, with results obtained when simulating a model with identical parameters in DigSILENT PowerFactory. The "Standard/Detailed model 2.2" synchronous machine model is chosen for all synchronous machines in PowerFactory. This is also a sixth order model, but the differential equations are formulated using fluxes instead of voltages. Saturation is neglected, and all the loads are constant impedance loads.


Figure \ref{fig:K2A-Validation} shows results from simulating a short circuit on Generator 1 in the Kundur Two-Area System, where the results from the Python simulation are shown alongside reference results from PowerFactory. The angle, speed and subtransient d- and q-axis voltages are shown. The results are very similar in this case, substantiating the validity of the model.

\section{Real-Time Simulation}
\label{sec:RealTimeSimulation}

\begin{figure}
    \centering
    \includegraphics{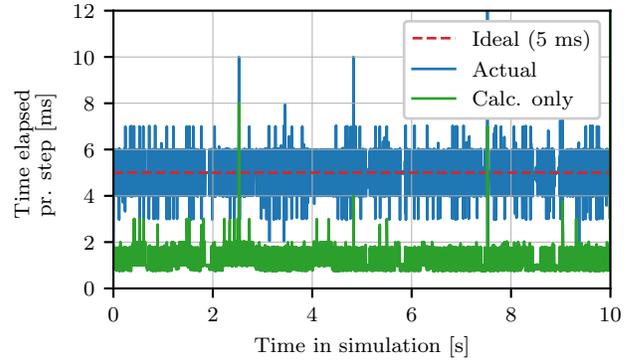}
    \caption{The figure indicates how the synchronization between exact time and simulation time is performed during real-time simulation of the IEEE 39-bus system, with a total of 123 states. The simulation step in this case is 5 ms (red, dashed line). The actual loop time varies somewhat around the ideal loop time (blue line). It can also be seen that the time spent only on the calculation of each step (green line) is on average around 1 to 2 ms, i.e. well below 5 ms.}
    \label{fig:RTSim-Synchronization}
\end{figure}

\begin{figure*}
    \centering
    \includegraphics[width=\textwidth]{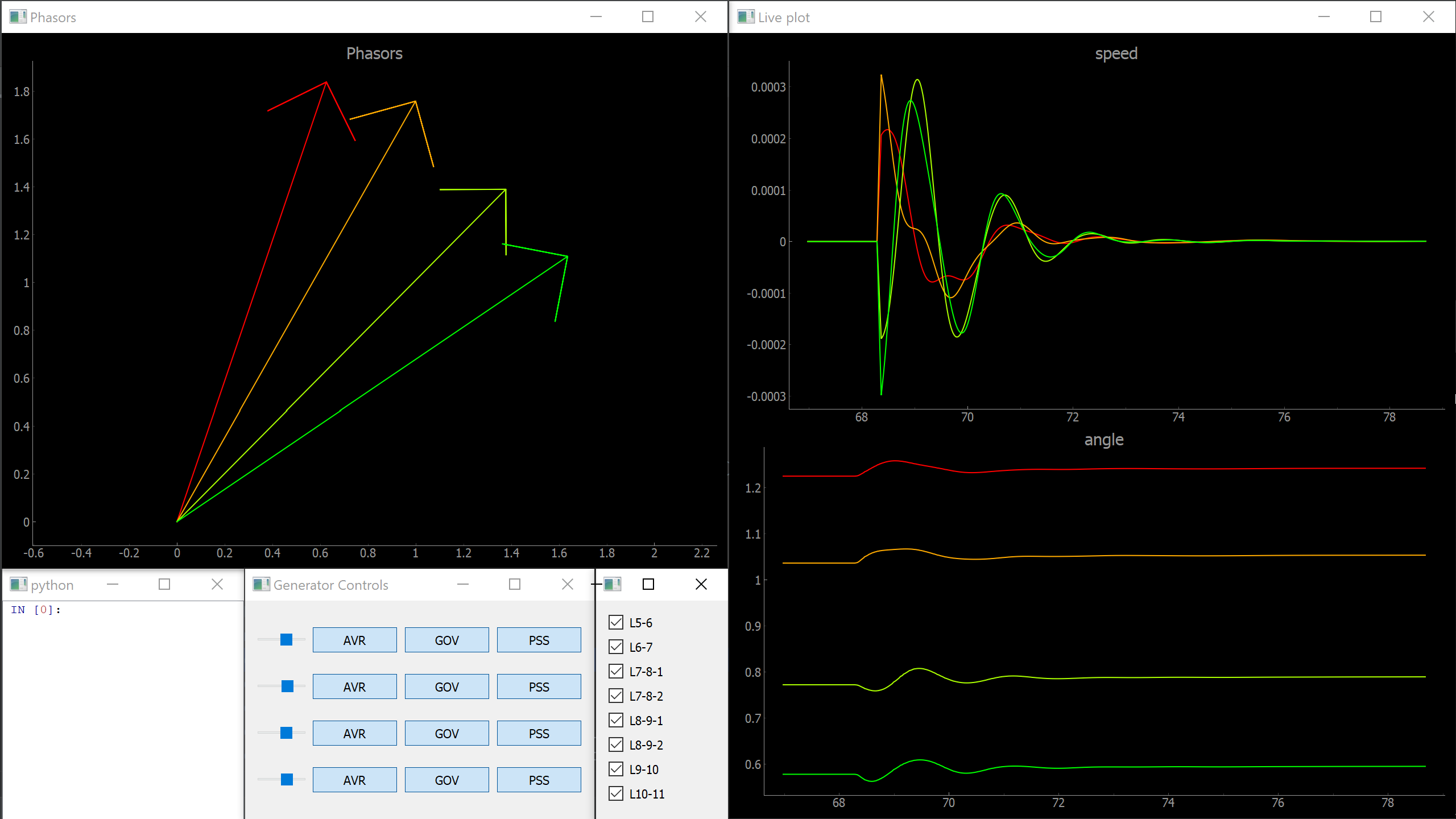}
    \caption{The GUI during real-time simulation of the Kundur Two-Area System is shown. The GUI is composed of a phasor plot displaying the rotor angle and excitation voltage magnitude ($E_f \angle \delta$) for each generator (upper left), and curves for generator power angles and speed (right). The controls (lower left) allows the user to interact with the simulation through a console (allowing commands to  be issued to modify the admittance matrix, generator control parameters etc.), activate/deactivate AVR, governor or PSS of any of the generators, or disconnect or connect one or more of the eight lines. If the AVR is deactivated, the excitation of the individual generators can be controlled manually by sliders.}
    \label{fig:GUI}
\end{figure*}

For smaller grids, the simulator is fast enough to perform real-time simulation with a time step of around 5 ms. To perform real-time simulation, a timer is started at the beginning of the simulation, which is used to synchronize the simulation time with the wall clock time. After each step taken by the solver, the simulator is set on hold until the next time step is due. Although Python is by no means a language tailored for real-time simulation, experience shows that the impression of continuous real-time interaction with the simulated model can be achieved.

Fig. \ref{fig:RTSim-Synchronization} shows the amount of time spent on each time step by the solver during real-time simulation of the IEEE 39-bus system with a time step of 5 ms. All generators are modelled using the sixth order model, and all generators except the largest synchronous machine (which represents the rest of USA and Canada) are equipped with the AVR, PSS and turbine-governor models presented in section \ref{sec:DynamicModels}. The model has a total of 123 states in this case, and is simulated using the Modified Euler method, with  constant step size and one correction iteration. Numba was used for Just-In-Time compilation of the state derivative functions to increase computational efficiency. The simulation was performed on a laptop with an Intel i7 CPU with two 2.8 GHz-cores and 16 GB RAM.

It can be seen that the calculation time (shown in green) on average is around 1 to 2 ms, which is fast enough for a time step of 5 ms (shown in red). The calculation time surpasses 5 ms in some occasions. Also, the actual time spent on each time step (shown in blue) varies approximately $\pm$2 ms around the ideal time step (with some more extreme outliers). Over time, the simulation time is relatively well synchronized with the real world time. It should be noted that the calculation time achieved here is orders of magnitude longer than what can be achieved with, for instance, DPsim \cite{Mirz2019}, which is specifically made to handle real-time simulation.

Most parameters of the power system model can be controlled directly during real-time simulation, and changes can be made to the admittance matrix. This makes it straight-forward to, for instance, apply short circuits, line outages etc. at will during the simulation, or change voltage references or active power set-points of generators.

For visualization and interaction with the real-time simulation, a simple Graphical User Interface (GUI) is implemented using the PyQtGraph \cite{Campagnola} library. This library allows custom GUIs to be implemented easily, with customized live plots and controls. A screen shot of the GUI is shown in Fig. \ref{fig:GUI} during real-time simulation of the Kundur Two-Area system. The GUI in this case shows live phasors representing the excitation voltage and rotor angle for each generator, as well as speed and angle plots. Additionally, the controls in the lower left part of the figure allows lines to be disconnected, and generator controls to be activated/deactivated.

\section{Discussion}
\label{sec:Discussion}

Based on the presented results, we elaborate on the following ideas:

\subsection{Desktop Real-Time Simulator}
It is found that real-time simulations of small to medium sized systems can be performed on an average laptop computer, without requiring specialized operative systems or hardware. For researchers, this potentially simplifies testing real-time capability and performing live demonstrations of WAMPAC applications significantly. In the educational setting, interactive real-time simulation and visualization allows students to get hands on experience with a running simulation model, which is valuable for developing intuitive understanding of different stability problems, control systems etc.

\subsection{Reproducibilty of research}
For researchers, it is often more time consuming than necessary to reproduce results from other simulation studies. Even small power system models have hundreds of parameters that need to be correct for researchers to be able to accurately reproduce results. Furthermore, the specific implementation of dynamic models might differ across simulation tools. If the entire code for reproducing results for a particular study could be made available for the community, complete with the power system model, dynamic models and the simulation code, this could significantly reduce the time spent on establishing the reference case. 

The presented simulation code being developed entirely in Python (relying only on standard packages for the core functionality) comes with the advantage that it is straightforward to upload the code to an online Python interpreter. This makes it easy to share, edit and run simulations in the cloud. This could make reproducing results even easier, as the simulation required to produce the published result could be performed in the internet browser.





\section{Conclusion}
Dynamic RMS simulations of small to medium sized grids can be performed efficiently in Python using the presented simulation software. This enables tight integration of power system simulations with other open source packages available for Python, which is the main objective. Although transparency and expandability is prioritized over computational efficiency and performance, it is found that real-time simulation with a reasonably short time step is possible. The possibility of uploading the entire code required for reproducing results from a publication is also emphasized.

Finally, adding that the package is free, easily deployable and cross-platform compatible, we believe that it could be a valuable tool, both for researchers prototyping WAMPAC applications, and in the educational setting.

\label{sec:Conclusion}
